\documentstyle[aps,twocolumn]{revtex}
\begin{document}
\draft
\title{Localization and fluctuations of local spectral density on
tree-like structures with large connectivity: Application to the
quasiparticle line shape in quantum dots}
\author{  Alexander D. Mirlin$^1\dagger$ and Yan V. Fyodorov$^2\dagger$}
\address{
$^1$ Institut f\"{u}r Theorie der Kondensierten Materie,
  Universit\"{a}t Karlsruhe, 76128 Karlsruhe, Germany}
\address{ $^2$ Fachbereich Physik, Universit\"at-GH Essen, 45117 Essen,
Germany}
\date{\today}
\maketitle
\tighten
\begin{abstract}
We study  fluctuations of the local density of states (LDOS)
on a tree-like lattice with large branching number $m$. The average
form of the local spectral function (at given value of the random
potential in the observation point) shows a crossover from the
Lorentzian to semicircular form at $\alpha\sim 1/m$, where $\alpha=
(V/W)^2$, $V$ is the typical value of the hopping matrix element, and $W$
is the width of the distribution of random site energies.
For $\alpha>1/m^2$ the LDOS fluctuations (with respect to this average
form) are weak. In the opposite case, $\alpha<1/m^2$, the fluctuations
get strong and the average LDOS ceases to be representative, which is
related to the existence of the Anderson transition at $\alpha_c\sim
1/(m^2\log^2m)$. On the localized side of the transition the spectrum
is discrete, and LDOS is given by a set of $\delta$-like peaks. 
The effective number of components in this regime is given by $1/P$,
with $P$ being the inverse participation ratio. It is shown that $P$
has in the transition point a limiting value $P_c$ close to unity,
$1-P_c\sim 1/\log m$, so that the system undergoes a transition
directly from the 
deeply localized to extended phase. On the side of delocalized states,
the peaks in LDOS get broadened, with a width
$\sim\exp\{-\mbox{const}\log m[(\alpha-\alpha_c)/\alpha_c]^{-1/2}\}$
being exponentially small near the transition point.
We discuss application of our results to the problem of the
quasiparticle line shape in a finite Fermi system, as suggested recently
by Altshuler, Gefen, Kamenev, and Levitov.

\end{abstract}
\narrowtext

\section{Introduction.}
\label{s1}

The recent experimental observation of single-level peaks in excitation
spectra of quantum dots \cite{sivan,sia} has motivated theoretical
interest in the problem of the quasiparticle life time induced by the
Coulomb interaction in mesoscopic samples \cite{sia,blanter}.
It was found that the single particle levels with excitation energy
$E<E_T$ have widths less than the one-particle mean level spacing
$\Delta$, and thus can be resolved. Here $E_T$ is the Thouless energy,
which has the physical meaning of the inverse time of spreading of a
wave packet over the system, the ratio $g=E_T/\Delta$ being the
dimensionless conductance of the dot. Thus, the theory predicts that
approximately the first $g$ single-particle levels can be resolved, in
agreement with experimental findings \cite{sivan,sia,ralph}.
In a more recent paper \cite{agkl}, Altshuler, Gefen, Kamenev and Levitov
(AGKL) mapped the problem onto a tree-like tight-binding model in the Fock
space. This model is very close to the Anderson tight-binding model on
the Bethe lattice, which was found \cite{abou,kunz,MF91} to undergo
the localization transition. This allowed AGKL to conclude that
there is a localization threshold $E_c$ ($\Delta < E_c < E_T$)
in the problem, so that the states well below $E_c$ are strongly
localized, i.e. given by the bare single-particle states with small
perturbative admixture of many-particle states.

The qualitative arguments used by AGKL are not sufficient, however, for
a more precise determination of the behavior of the spectral shape of the
quasiparticle peak in the vicinity of the localization threshold.
On the other hand, the critical behavior of the Anderson model on the
Bethe lattice was studied in detail in \cite{MF91}. The critical
behavior found there is completely analogous to that obtained earlier
for the  Bethe lattice version of the $\sigma$-models
\cite{Efe-bl,Zirn-bl} which
can be derived from Wegner's $n$-orbital model with $n\gg 1$
states per lattice site \cite{Weg} (or, equivalently, considering a
system of weakly coupled metallic granules \cite{Efe85}).
Exactly the same type of the transition and critical behavior
on the Bethe lattice has been found \cite{gruz}
in the ``toy'' version of the supersymmetric $\sigma$-model introduced
in \cite{hyper}, which is much simpler technically than the ``true''
$\sigma$-models of localization. In Ref.\cite{ldos-bethe} fluctuations
of the local density of states (LDOS) near the Anderson transition
were studied; as we will see, this question is closely related
to the subject under discussion.

The aim of this paper is to study the model derived by AGKL via the
supersymmetry technique. In fact, since many results for the critical
behavior on the Bethe lattice are already known, we have just to apply
them to the present model, properly identifying parameters and
quantities of interest. There are, however, new features, which
require an additional investigation. First of all, the model of AGKL
has a large connectivity (branching number) $m\gg 1$. Secondly,
we will be  interested in the structure of the local spectral function
in energy space; the question, which was not addressed in
Refs. \cite{MF91,ldos-bethe}.

The outline of the paper is as follows. In Sec.\ref{s2} we formulate
the model as derived by AGKL and identify the parameters and
quantities to be studied. Sec.\ref{s4} is devoted to calculation of the
average shape of
the local spectral function on a Bethe lattice with large branching
number. In Sec.\ref{s4a} we evaluate the magnitude of the LDOS
fluctuations in the region where these
fluctuations are weak.
In Sec.\ref{s3} we study the correlations of
amplitudes of eigenstates on a tree-like lattice in the critical
region. This information is used
 in Sec.\ref{s5}, where we study fluctuations of the local
spectral function and its typical shape in the vicinity of the
Anderson transition, both in the phases of localized and extended states.
In Sec.\ref{s5a} we discuss application of this results to the problem
of quasiparticle excitations in quantum dots.
Our results are summarized in Sec.\ref{s6}.

\section{Formulation of the problem.}
\label{s2}

For the sake of completeness and clarity, we recall the derivation of
the model by AGKL \cite{agkl}. To describe the system of interacting
fermions one should study eigenstates $|\alpha\rangle$ of a generic many-body
Hamiltonian of $N$ pairwise interacting particles :
\begin{equation}\label{genham}
{\cal H}=\sum_{p}{\cal E}_p a_p^{\dagger}a_p+\sum_{p,q,r,s}V^{p,q}_{r,s}
a_p^{\dagger}a_q^{\dagger}a_ra_s
\end{equation}
where ${\cal E}_p $ are energies of the corresponding single particle
states. When one neglects the interaction $\hat{V}$ the complete
set of eigenstates is provided by Slater determinants
obtained from the filled 
Fermi sea $|0_N\rangle$ as
\begin{equation}\label{slat}
|\Psi_N^{\{p\}}\rangle=a_{p_{2m}}^{\dagger}...a_{p_{m+1}}^{\dagger}
a_{p_m}...a_{p_1}|0_N\rangle
\end{equation}
As a result of interaction, the exact eigenstates $|\alpha\rangle$ for
interacting particles are superpositions of the Slater determinants
$|\alpha\rangle=\sum_{\{p\}}
A_{\{p\}}^{\alpha}|\Psi_N^{\{p\}}\rangle$.

To analyze the form of the line for single-particle excitations
AGKL suggested to consider the
hierarchy of the
many-body states in the Fock space. Let us consider the system with
$N$ particles in its Hartree-Fock ground state $|\tilde{0}_N\rangle$.
The first level of hierarchy
is formed by states with one particle:
$|1_p\rangle=\tilde{a}_p^{\dagger} 
|\tilde{0}_N\rangle$, second by the states with
two particles and one hole (which we will denote as
three-particle states for brevity):
$|3_{qpr}\rangle = 
\tilde{a}_q^{\dagger}\tilde{a}_p^{\dagger}\tilde{a}_r|\tilde{0}_N\rangle$, 
the third one by the states with
three particles and two holes (to be denoted as
five-particle states) and so forth.
Neither of these states ( which we denote generally as $|i\rangle$ )
is an eigenstate of ${\cal H}$. Therefore
if an extra particle is added to the system in such a way
that  a single-particle state $|1\rangle$
on the first level of hierarchy is formed, this state will decay
 to the states on the second level of hierarchy, which in turn may
spread further in the Fock space.
The process of the spread may be
looked at as a quantum diffusion of a fictitious particle
on a graph whose vertices $i$ are formed by the states $|i\rangle$
at different levels of
hierarchy. As such, it can be described by an effective
one-particle tight-binding
model characterized by the Hamiltonian:
\begin{equation}
\hat{H}=\sum_i\epsilon_i c_i^\dagger c_i+  \sum_{\langle ij\rangle}
V_{ij} c_i^\dagger c_j\ ,
\label{e1a}
\end{equation}
with "site energies" $\epsilon_i=\langle i|{\cal H}|i\rangle$ and
hopping constants $V_{ij}=\langle i|{\cal H}|j\rangle$.

Let us note that a problem of decay of an isolated level to a dense band
in a situation when the level is directly coupled only to a small fraction
of the states of the continuum was studied by Akulin and Dykhne \cite{Akulin}. Though the model
considered in \cite{Akulin} is somewhat different from that studied in the present paper, many physical features are actually similar for both models.
In particular, it was stressed in \cite{Akulin} that the dynamics of level decay can by affected by the phenomenon of Anderson localization.

The spectral shape of a quasiparticle peak is obtained by projecting the
initial single-particle state
$|1\rangle$ onto exact eigenstates $|\alpha\rangle$ \cite{agkl}:
\begin{equation}
\rho_1(E)=\sum_\alpha |\langle\alpha|1\rangle|^2 \delta(E-E_\alpha)
\label{e2}
\end{equation}
If we now consider the state $|1\rangle$ as a site of the
tight-binding model, we see that $\rho_1(E)$ is nothing else but the
local density of states (LDOS) in the point $1$ at energy $E$, i.e.
the Fourier-transform
\begin{equation}
\rho_1(E)={1\over 2\pi}\int_{-\infty}^{\infty}dte^{iEt}\left\langle  
1\left|e^{-i\hat{H}t}\right|1\right\rangle
\end{equation}
of the overlap of the spreading wave function with the initial
state  $|1\rangle$.
In another context this quantity is called the strength
function and is frequently used in nuclear and atomic
physics to characterize spectra of complex
systems,  see e.g. \cite{Grib} for recent applications
and further information.

The interaction
matrix elements couple the states of each generation to those of the
preceding and of the following generations. In particular, a
single-particle state is connected with three-particles
states, but not with five-, seven-, \ldots -particle ones.
The density of one-particle states
is equal to $\nu_1=1/\Delta$.
Since we are interested in the regime,
where one-particle states can be resolved or, in other words, the
width is less than $\Delta$, we can keep only the states with energies
within a window ($E-\Delta/2, \ E+\Delta/2$) containing just one
single-particle state.
The density of three-particle states at
given energy $E\gg\Delta$ is given by
\begin{equation}
\nu_3(E)\simeq{1\over 2\Delta^3}\int_0^E dE_1 \int_0^{E_1} dE_2=
{E^2\over 4\Delta^3}\ ,
\label{e1}
\end{equation}
the factor $1/2$ taking care of the identity of the
quasiparticles.
The ratio $m=\nu_3(E)/\nu_1=E^2/ 4\Delta^2$ defines the effective
branching number of the problem and is considered as a big parameter
 \cite{note1}.
The important fact observed by AGKL is that this branching number
stays parametrically the same for the following levels of hierarchy.
For example, the density of the five-particle states $\nu_5(E)\sim
E^4/\Delta^5$ and each of these states is connected with of order of
one three-particle state (in the chosen energy window of the width
$\Delta$), whereas each three-particle state is connected to
$\sim E^2/\Delta^2\sim m$ five-particle states. Following AGKL, we
will neglect the change in numerical coefficients, and consider a tree
structure with constant branching number $m\gg 1$ (Bethe lattice)
\cite{note2}.

The Hamiltonian of the Anderson model on the Bethe lattice  is determined
essentially by the following three parameters: the branching number
$m$, the characteristic magnitude $V$ of the hopping matrix elements,
and the width $W$ of the distribution of the random site energies
$\epsilon_i$. They are related to the parameters of the original
(quantum dot) problem as follows. The connectivity $m$ is determined by
the proliferation of the density of states from one level of the
hierarchy to the next one, as discussed above:
\begin{equation}
m\sim\nu_3(E)/\nu_1\sim\nu_5(E)/\nu_3(E)\sim\ldots\sim E^2/\Delta^2
\label{e1b}
\end{equation}
The characteristic value of the hopping matrix element is equal to
\cite{blanter,agkl}
\begin{equation}
V\sim\Delta^2/E_T\sim\Delta/g
\label{e1c}
\end{equation}
Finally, $W$ can be determined by equating the mean spacing of the
states in the next level of the hierarchy connected to a given state,
$1/\nu_3(E)\sim\Delta^3/E^2$, to the corresponding quantity in the Anderson
model on the Bethe lattice, $W/m$. This yields:
\begin{equation}
W\sim m\Delta^3/E^2\sim\Delta
\label{e1d}
\end{equation}
In order to check that restriction to the states in the energy interval
$\Delta$ is justified, we can modify the model by including all states
from a broader energy window $W_n=n\Delta$, with $1\ll n\ll
E/\Delta$. Then each three-particle state will be coupled to $n$
one-particle states; each five-particle state to of order of $n$
three-particle states etc. This modified model thus resembles the
$n$-orbital model on the Bethe lattice. It has branching number  given
by the same expression (\ref{e1b}) and the coupling constant
\cite{Weg,Efe85} 
$\alpha\sim(nV/W_n)^2=(V/\Delta)^2\sim 1/g^2$, independent of $n$,
which confirms that $n$ is indeed an irrelevant parameter. 
Equivalence of the $n$-orbital and the Anderson ($n=1$) models
at $\alpha\ll 1$ will be discussed in more detail elsewhere \cite{tobe}.

It is well known \cite{abou,kunz,MF91} that the Anderson model on the Bethe lattice exhibits the Anderson localization transition when one of control
parameters $W,V,m$ varies. In our case, the change in the quasiparticle excitation energy $E$ implies, see eq.(\ref{e1b}) , the change of 
the branching number $m$. Such a phase transition is reflected in a drastic
change of statistical properties of eigenfunctions.
In our preceding publications
\cite{MF91,ldos-bethe} we have discussed the distribution function of
LDOS and its typical spatial shape (for given energy)
near the Anderson transition.
Now, our interest is concentrated on a typical shape of $\rho_1(E)$
in energy space, for a given ``point'' $|1\rangle$.

\section{Average local spectral function.}
\label{s4}

In this section, we study the average form of the local spectral
density $\rho_1(E)$ in a site $|1\rangle$ of a Bethe lattice with
large connectivity (branching number) $m\gg 1$. We consider for
definiteness a model with site-diagonal disorder
\begin{equation}
\hat{H}=\sum_i\epsilon_i c_i^\dagger c_i+ V \sum_{\langle ij\rangle}
c_i^\dagger c_j
\label{e10}
\end{equation}
The averaging is performed over the random energies $\epsilon_i$ in
all sites except the site $|1\rangle$, where the spectral density is
studied. Thus, we calculate the average form of the LDOS in a site
with given value $\epsilon_1$ of the random potential. The same object
was studied previously in a random banded matrix  model with
strongly fluctuating  diagonal elements \cite{SRBM}. We will compare
the results for the present case to those of the Ref.\cite{SRBM} in
the end of the section.

The average value of the LDOS is given by:
\cite{MF91}
\begin{equation}
\rho_1(E)={1\over\pi}\mbox{Re}\int dR\,R
F^{m+1}(R^2)e^{i(E-\epsilon_1)R^2/2}\ ,
\label{e11}
\end{equation}
where $F(R^2)$ is the solution to the following non-linear integral equation:
\begin{eqnarray}
&&F(S^2)=1-S\int_0^\infty dR\int d\epsilon \nonumber\\
&& \times \gamma(\epsilon)
e^{i(E-\epsilon)R^2/2} F^m(R^2) V J_1(VRS) \label{e12}
\end{eqnarray}
Here $\gamma(\epsilon)$ is the distribution of the random site
energies $\epsilon_i$, and $J_1(x)$ is the Bessel function. We look
for an approximate solution of eq.(\ref{e12}) in a Gaussian form
\begin{equation}
F(R^2)=\exp\{- g_0 R^2\}
\label{e13}
\end{equation}
Substituting eq.(\ref{e13}) in the r.h.s. of eq.(\ref{e12}), we get
\begin{eqnarray}
&&\mbox{r.h.s.(\ref{e12})}= \int d\epsilon\gamma(\epsilon)
\exp\left\{-{1\over 4} V^2S^2{1\over mg_0-i(E-\epsilon)/2}\right\}
\nonumber \\
&&=\exp\left\{-{1\over 4} V^2S^2 \int d\epsilon\gamma(\epsilon)
{1\over mg_0-i(E-\epsilon)/2}\right\} \nonumber\\
&& \times [1+g_1 S^4+\ldots]\ ,
\label{e14}
\end{eqnarray}
where
\begin{eqnarray}
g_1&=&{1\over 32}V^4\left\{
\int d\epsilon\gamma(\epsilon) {1\over [mg_0-i(E-\epsilon)/2]^2}
\right. \nonumber
\\
&-&\left.\left[\int d\epsilon\gamma(\epsilon) {1\over
mg_0-i(E-\epsilon)/2}\right]^2
\right\}
\label{e15}
\end{eqnarray}
~From eqs.(\ref{e12}), (\ref{e13}), and (\ref{e14}) we get an equation
for $g_0$:
\begin{equation}
g_0={V^2\over 4}
\int d\epsilon\gamma(\epsilon) {1\over mg_0-i(E-\epsilon)/2}
\label{e16}
\end{equation}
The Ansatz (\ref{e13}) is justified if the $O(S^4)$ correction in
eq.(\ref{e14}) can be neglected. Since
\begin{equation}
F^m(R^2)=\exp\{-mg_0R^2\}(1+mg_1R^4+\ldots)\ ,
\label{e17}
\end{equation}
the integrals in the r.h.s. of eqs.(\ref{e11}), (\ref{e12}) are
dominated by the region $R^2\sim 1/mg_0$. Furthermore, according to
eq.(\ref{e15}), $g_1\sim g_0^2$. Thus, the $R^4$ term in
eq.(\ref{e17}) leads to a relative correction of order of $1/m$, which
is small in the case of large connectivity, $m\gg 1$.  Therefore,
eq.(\ref{e13}) is indeed the proper approximation for $m\gg 1$.

Now we will analyze eq.(\ref{e16}) in the two regimes: (i) $(V/W)^2\ll
1/m$, and (ii) $(V/W)^2\gg 1/m$, where $W$ is a typical magnitude of
the random energies $\epsilon_i$. In the case (i) one can neglect
$g_0$ in denominator of (\ref{e16}). This yields
\begin{eqnarray}
\mbox{Re} g_0 &=& {\pi\over 2} V^2\gamma(E)\equiv {v\over 2}
\nonumber\\
\mbox{Im} g_0 &=& {1\over 2}V^2\int d\epsilon\gamma(\epsilon){1\over
E-\epsilon} \equiv {u\over 2}
\label{e18}
\end{eqnarray}
Since a typical scale for $\gamma(E)$ is $\gamma(E)\sim W^{-1}$
we find $g_0\sim V^2/W$. We see that under the condition $(V/W)^2\ll
1/m$ the quantity $mg_0$ can be indeed neglected in
the denominator of eq.(\ref{e16})
 as compared to $E-\epsilon\sim W$. Then the LDOS,
eq.(\ref{e11}), takes the Breit--Wigner form
\begin{eqnarray}
\langle \rho_1(E)\rangle &=& {1\over 2\pi}\mbox{Re}{1\over
(m+1)g_0-i(E-\epsilon_1)/2} \nonumber\\
&\simeq& {1\over \pi}{mv\over (mv)^2+(E-\epsilon_1-u)^2}
\label{e19}
\end{eqnarray}
(we neglected the difference between $m+1$ and $m$ in the second
line), i.e. it is a Lorentzian with a width
\begin{equation}
\Gamma/2=mv=m\pi V^2\gamma(E)\sim mV^2/W\ll W
\label{e20}
\end{equation}
The center of this Lorentzian is determined by the local random energy
$\epsilon_1$ (slightly shifted by $u\sim\Gamma$).
The width $\Gamma$ is exactly the quasiparticle decay rate one would
get by applying the Fermi Golden Rule: it is equal to $2\pi$ times
matrix element $V$ squared times density of states into which the
particle decays ($m\gamma(\epsilon_1)\simeq m\gamma(E)$).

In the opposite case, $(V/W)^2\gg 1/m$, we can neglect $\epsilon$ in the
denominator of eq.(\ref{e16}), which reduces then to
\begin{equation}
g_0={V^2\over 4}{1\over mg_0-iE/2}
\label{e21}
\end{equation}
We find therefore
\begin{equation}
g_0={1\over 4m}(iE+\sqrt{4mV^2-E^2})
\label{e22}
\end{equation}
and
\begin{eqnarray}
&&\langle\rho_1(E)\rangle={1\over 2\pi}\mbox{Re}{1\over mg_0-iE/2} =
{2\over \pi V^2}\mbox{Re} g_0 \nonumber\\
&&=\left\{
\begin{array}{ll}
{1\over 2\pi m V^2}\sqrt{4mV^2-E^2}\ , & \qquad |E|\le 2\sqrt{m}V\ ,\\
0\ , & \qquad |E|\ge 2\sqrt{m}V
\end{array}
\right. \label{e23}
\end{eqnarray}
Thus, the LDOS in this limit has a semicircular form. The width
of the semicircle is of order of $\sqrt{m}V\gg W$
justifying omission of $\epsilon\lesssim W$
in the denominator of eq.(\ref{e16}).

The crossover from the Lorentzian to the semicircular shape of local
DOS is completely analogous to that found for the model of
random banded matrices with strongly fluctuating diagonal elements
\cite{SRBM}. The physical reason for this crossover is the same in
both cases. When the disorder is sufficiently strong, an eigenstate is
not spread uniformly over all the sites of the lattice,
but rather is concentrated on a small fraction of lattice, formed by
the sites which satisfy a kind of resonance condition with a given
site $|1\rangle$. This leads to a Lorentzian form of the LDOS with a
width $\Gamma$ much less than the width $W$ of the distribution of
diagonal matrix elements (random energies). Position of this
Lorentzian for a site $|i\rangle$
is determined essentially by the random energy $\epsilon_i$.
The weaker is the disorder the larger is $\Gamma$. The crossover point is
determined by the condition $\Gamma\sim W$. For a still weaker
disorder the LDOS acquires the semicircular form, one and the same for all
sites of the lattice. In this regime eigenstates are spread over all
the lattice like in the Gaussian ensemble.

Finally, we note that when the Bethe lattice
model is used to describe excitations in a quantum dot only the
first (Lorentzian) regime is relevant. Indeed, in this case one is
interested in the regime $\Gamma<\Delta$ when the single-particle
levels are resolved, whereas the energies $\epsilon_i$
of many-particle states included in the Bethe lattice model
are distributed in the interval of the order of
$\Delta$, see Eq.(\ref{e1d}). The Lorentzian form of the local spectral
function holds, of course, also for higher energies, $E\gtrsim E_T$,
where the discreteness of the spectrum is unimportant and the Fermi
Golden Rule is certainly applicable. To see this in our approach,
one has to include in the model the states from a broader energy
interval $n\Delta>\Gamma$, as discussed in the end of
Sec.\ref{s2}. However, we do not consider this region, where the
single-particle states are not resolved and all the results (say, for
the lifetimes) can be simply obtained in perturbation theory.

\section{Fluctuations of LDOS: regime of weak fluctuations}
\label{s4a}

In the preceding section, we have calculated the local spectral
function $\rho_1(E)$ at fixed value of the random potential
$\epsilon_1$ in a given site $|1\rangle$ but averaged over all the
other random energies $\epsilon_i$, $i\ne 1$. The main aim of the
remaining part of the paper
is to find out how a typical (rather than averaged) LDOS looks
like. For this purpose, we study fluctuations of $\rho_1(E)$ in
various regimes. It turns out that the parameter
 which distinguishes between the
regimes (at fixed value of the branching number $m$ \cite{note4}) is
\begin{equation}
\alpha\sim (V/W)^2
\label{e30}
\end{equation}
Our consideration in this section will be similar to that of the
preceding one. This will allow us to describe fluctuations of the LDOS
in the regime where they are weak, namely $\alpha\gg 1/m^2$.

To study the fluctuations of the LDOS, it is not sufficient to
consider the integral equation (\ref{e12}) for the function $F(S^2)$
determining the average LDOS. Instead a more complicated equation on a
function $F(S_1^2,S_2^2)$ of two variables has to be considered
\cite{MF91}:
\begin{eqnarray}
&&F(S_1^2,S_2^2)=4\int R_1 dR_1\int R_2 dR_2 J_0(VR_1S_1)J_0(VR_2S_2)
\nonumber\\
&&\times{\partial^2\over\partial(R_1^2)\partial(R_2^2)} 
 F^m(R_1^2,R_2^2)\int d\epsilon\gamma(\epsilon)\nonumber\\
&& \times
\exp[i(E-\epsilon)(R_1^2-R_2^2)/2]  \label{e60}
\end{eqnarray}
Let us look again for its approximate solution in the Gaussian form:
\begin{equation}
F(S_1^2,S_2^2)=\exp\left\{-{i\over 2}S_1^2(u-iv)+{i\over 2}S_2^2(u+iv)
\right\}
\label{e61}
\end{equation}
Substituting (\ref{e61}) in the r.h.s. of eq.(\ref{e60}), we get
\begin{eqnarray}
&&\mbox{r.h.s.(\ref{e60})} = \int d\epsilon\gamma(\epsilon)\exp\left\{
{V^2S_1^2\over 2i}{1\over -m(u-iv)+E-\epsilon} \right.\nonumber\\ 
&&\left.
+{V^2S_2^2\over 2i}{1\over m(u+iv)-(E-\epsilon)}
\right\}
\label{e62}
\end{eqnarray}
Expanding this expression up to the terms of
the order of $S_1^2$ and $S_2^2$ one finds that
 $(v+iu)/2\equiv g_0$  satisfies the same equation
(\ref{e16}) obtained in sec.\ref{s4} where the average LDOS was
considered.

The function $F(S_1^2,S_2^2)$ determines the fluctuations of the
one-site Green's function
$$
G_1(E)=\sum_\alpha|\langle\alpha|1\rangle|^2(E-E_\alpha+i0)^{-1}
$$
(imaginary part of which is equal to $-\pi\rho_1(E)$) via the
following equation \cite{ldos-bethe}
\begin{eqnarray}
&& F^{m+1}(R_1^2,R_2^2)\exp\{i(E-\epsilon_1)(R_1^2-R_2^2)/2\}
\nonumber\\
&&=\int_{-\infty}^{\infty}d\tilde{u}\int_{0}^{\infty}d\tilde{v}
{\cal P}(\tilde{u},\tilde{v})\exp\{i\tilde{u}(R_1^2-R_2^2)-
\tilde{v}(R_1^2+R_2^2)\} \nonumber \\ &&
\label{e63}
\end{eqnarray}
Here $\tilde{u}$ and $\tilde{v}$ are the real and imaginary part of
the inverse Green's function,
\begin{equation}
[G_1(E)]^{-1}=\tilde{u}-i\tilde{v}\ ,
\label{e63a}
\end{equation}
and ${\cal P}(\tilde{u},\tilde{v})$ is the joint probability
distribution of $\tilde{u}$ and $\tilde{v}$.

Substituting eq.(\ref{e61}) into (\ref{e63}), we find
\begin{equation}
{\cal P}(\tilde{u},\tilde{v})=
\delta(\tilde{u}-\tilde{u}_0)\delta(\tilde{v}-\tilde{v}_0)\ ,
\label{e64}
\end{equation}
where (neglecting the difference between $m+1$ and $m$ as in
Sec.\ref{s3})
\begin{eqnarray}
\tilde{u}_0&=&{1\over 2}(E-\epsilon_1-mu)\ ; \label{e65}\\
\tilde{v}_0&=&{mv\over 2} \label{e66}
\end{eqnarray}
We see that eqs.(\ref{e65}), (\ref{e66}) together with the definition
(\ref{e63a}) reproduce precisely the average LDOS found in the
preceding section. Therefore, the Gaussian
approximation (\ref{e61}) corresponds
 to absence  of any LDOS fluctuations, as is clearly seen from
eq.(\ref{e64}). To calculate corrections to the Gaussian
approximation we make the next iteration  and expand eq.(\ref{e62})
up to $O(S^4)$ terms. Assuming that we are in the Lorentzian
regime, $(V/W)^2\ll 1/m$, we get after some algebra
\begin{eqnarray}
F(S_1^2,S_2^2)&=&\exp\left\{-{i\over 2}S_1^2(u-iv)+{i\over 2}S_2^2(u+iv)
 \right\}\nonumber\\
&\times&\left[1+{\pi\over 4}{\gamma(E-mu)\over mv} V^4S_1^2 S_2^2
 +\ldots\right]
\label{e67}
\end{eqnarray}
Substituting this in the l.h.s. of eq.(\ref{e63}) and expanding both
sides of this equation in a power series in $(R_1^2-R_2^2)$ and
$(R_1^2+R_2^2)$, we find the average values
$\langle\tilde{v}\rangle=\tilde{v}_0$,
$\langle\tilde{u}\rangle=\tilde{u}_0$
(as given by eqs.(\ref{e65}), (\ref{e66})), and the variances
\begin{equation}
\mbox{var}(\tilde{v})=\mbox{var}(\tilde{u})={\pi\over 8}V^4
{\gamma(E-mu)\over v}\simeq {V^2\over 8}
\label{e68}
\end{equation}
The parameters $\tilde{u}_0$ and $\tilde{v}_0$ determine the position
and the width of the averaged (Lorentzian) LDOS whereas
eq.(\ref{e68}) describes its fluctuations. The relative strength of
the fluctuations is characterized by the ratio
\begin{equation}
{ \mbox{var}(\tilde{v}) \over \langle\tilde{v}\rangle^2 }=
{\pi\over 4}V^3 {\gamma(E-mu)\over v^2}\simeq{1\over 4m\pi\gamma(E)V}
\sim {W\over mV}
\label{e69}
\end{equation}
Thus, the LDOS fluctuations are weak when $\alpha\gg 1/m^2$. In the
opposite case the fluctuations are strong and the Gaussian Ansatz
is not a good approximation any longer.
Correspondingly, the expansion around it employed in this section
breaks down.

\section{Eigenfunctions correlations near the localization threshold}
\label{s3}

Now we want to study a typical structure of LDOS in the vicinity of the
Anderson transition point. For this purpose, we will need, however,
 information concerning the  correlations in amplitudes of two different
eigenfunctions with energies close to the mobility edge
in the same spatial point.
We will not use the fact that $m\gg 1$ here and so our calculation will be valid
for arbitrary branching number. Moreover, we believe that the results
of this section are of even more general validity and are not
restricted to tree-like lattices (see the discussion in the end of the
section).

The quantity we want to calculate here is the correlation function
$\sigma(r,E,\omega)$ representing the overlap of the eigenfunctions,
\begin{eqnarray}
\sigma(r,E,\omega)&=&\Delta^{2}\tilde{R}_2^{-1}(\omega)
\langle\sum_{\alpha\ne\beta}|\Psi_\alpha(r)|^2|\Psi_\beta(r)|^2 \nonumber\\
&\times&\delta(E-E_\alpha)\delta(E+\omega-E_\beta)\rangle\ ,
\label{e3}
\end{eqnarray}
where $\tilde{R}_2(\omega)$ is the level correlation function,
\begin{equation}
\tilde{R}(\omega)=\Delta^{2}
\langle\sum_{\alpha\ne\beta}
\delta(E-E_\alpha)\delta(E+\omega-E_\beta)\rangle\ .
\label{e3a}
\end{equation}
We are going to study it for energies $E$, $E+\omega$
 close to the mobility edge (in the
phase of delocalized states) so that the correlation length $\xi$
is large. We will assume however that the system size is larger than
the correlation length, so that the system is in the critical regime,
but not exactly in the critical point. We will compare
$\sigma(r,E,\omega)$
 to the analogous correlation function containing a single eigenfunction,
\begin{equation}
\pi(r,E)=\Delta \langle \sum_\alpha |\Psi_\alpha(r)|^4
\delta(E-E_\alpha)\rangle\ ,
\label{e4}
\end{equation}
which defines the inverse participation ratio,
\begin{equation}
P(E)=\sum_r \pi(r,E)
\label{e4a}
\end{equation}

In fact, the Bethe lattice is not a proper model to study quantities
like $\pi(r,E)$ since these quantities are well-defined
only for finite systems. However, a finite Bethe lattice
 has a peculiar feature: most of its sites
 are located at the boundary. As the result, the properties
may depend crucially on boundary conditions even in the thermodynamic
limit.
There exists, however, another model -- that of sparse
random matrices (SRM) --  which describes a lattice having locally a
tree-like structure similar to that of the Bethe lattice but
possessing also large-scale loops ensuring that all sites are essentially
equivalent. As a consequence, this model is free from the problem of
boundary conditions. At the same time it is known that
 the localization transition in the SRM
model is equivalent to that on the Bethe lattice \cite{sparse}.
The inverse participation ration (\ref{e4a}) was studied in the
delocalized phase of the SRM model near the localization threshold in
Ref.\cite{sparse-prl}.
It was shown there that its critical behavior is determined by an
exponentially large scale $C(E)$ emerging in the non-compact sector
of the effective supersymmetric model,
\begin{eqnarray}
&& P(E)\propto N^{-1}C(E) \label{e4c}\\
&& C(E)\propto\exp\{\mbox{const}|E-E_c|^{-1/2}\}
\label{e4b}
\end{eqnarray}
The scale $C(E)$ has a physical meaning of the ``correlation volume''
\cite{ldos-bethe} and determines also the critical behavior of
conductivity in the delocalized phase.
We extend now the calculation of Ref.\cite{sparse-prl} to find also the
correlation function (\ref{e3}).


To calculate the overlap function defined in Eq.(\ref{e3})
we  follow \cite{corr} and use the identity relating
$\pi(r,E)$ and $\sigma(r,E,\omega)$ to
 advanced and retarded Green functions $G^{R,A}(
r,E)=\sum_{\alpha=1}^N \left|\Psi_\alpha(r)\right|^2 (E\pm
i0-E_\alpha)^{-1}$:
\begin{eqnarray}
&& 2\pi^2\left[\Delta^{-1}\pi(r,E)\delta(\omega)+\Delta^{-2}
\tilde{R}_2(\omega)\sigma(r,E,\omega)\right] \label{4}\\
&&= \mbox{Re}\left[\langle
G^R(r,E)G^A(r,E+\omega)  -
 G^R(r,E)G^R(r,E+\omega)\rangle \right] \nonumber
\end{eqnarray}
 Let us consider for definiteness the ensemble of real symmetric
SRM, corresponding to systems with unbroken time-reversal invariance.
For any site index $r=1,...,N$ we introduce one eight-component
supervector $\Phi^{\dagger}=\left(\Phi^{\dagger}_R,
\Phi^{\dagger}_A\right)$  consisting of two
four-component supervectors
$\Phi_{\sigma}^{\dagger}=\left(\phi_{\sigma,b1},\phi_{\sigma,b2},
\phi_{\sigma,f}^*,-\phi_{\sigma,f}\right)$, where indices $\sigma=R,A$
and $b,f$ are used to label advanced-retarded and boson-fermion
subspaces, respectively.
The ensemble-averaged products $\langle G^{\sigma}G^{\sigma'}\rangle$
for RSM model in the limit $N\gg 1$
can be extracted from the paper \cite{sparse} and the
Appendix D of the paper\cite{FM-ema} and is given by:
\begin{eqnarray}
&& \left\langle
G^{\sigma}(r,E)G^{\sigma'}(r,E+\omega) \right\rangle=
\left(1-\frac{4}{3}\delta_{\sigma,\sigma'}\right) \nonumber
\\ \nonumber
&& \times
\int DQ\left\langle \phi_{\sigma,b1}\phi_{\sigma,b1}\phi_{\sigma',b1}
\phi_{\sigma',b1}\right\rangle_{g_T}
\exp{\left(\frac{i\pi\rho\omega N}{4}
\mbox{Str } Q \Lambda\right)}\ ;\\
&&\langle \ldots\rangle_{g_T} = \int
 d\Phi (\ldots) \exp{\left[\frac{i}{2}E\Phi^{\dagger}L\Phi+
mg_T(\Phi)\right]}.\label{5}
\end{eqnarray}
 The function $g_T(\Phi)\equiv g_0(\Phi^{\dagger}
T^{\dagger}T\Phi;\Phi^{\dagger}L\Phi)$ satisfies the integral equation:
\begin{equation}\label{6}
g_T(\Psi)=\left\langle \left[h_F\left(\Phi^{\dagger}L\Psi\right)-1\right]
\right\rangle_{g_T}\ ,
\end{equation}
where $h_F(t)=\int dz e^{-itz}h(z)$ is the Fourier-transform of the
distribution of nonzero elements of the SRM. The
$8\times 8$ supermatrices $T$ satisfy the condition $T^{\dagger}LT=L$
 where $L=\mbox{diag}(1,1,1,1,-1,-1,1,1)$
and belong to a graded coset space whose explicit parametrization
can be found in \cite{Efrev,VWZ}. The supermatrices $Q$ are expressed
in terms of $T$ as $Q=T^{-1}\Lambda T$.  At last, the matrix $\Lambda=
\mbox{diag}(1,1,1,1,-1,-1,-1,-1)$, and the density of states $\rho$
is expressed in terms of the solution of the equation Eq.(\ref{6})
as $\rho(E)=-2g_{0x}/(\pi B_2)$, where $B_2=\int dz h(z) z^2$
and $g_{0x}=\partial g_{0}(x,y)/\partial x|_{x,y=0};\quad
x=\Phi^{\dagger} \Phi,\quad y=\Phi^{\dagger}L \Phi$.

When deriving  Eq.(\ref{5}), evaluation of a functional
integral by the saddle-point method has been employed, see details in
\cite{sparse,FM-ema}. An accurate consideration shows that such a
procedure
is legitimate as long as: i) the matrix size $N$ (playing in our model
the role of the volume) is large enough (much larger than the
coefficient $C(E)$
determining the size dependence of IPR, see above); and
ii) the energy difference $\omega$ is small enough (much smaller than
$C^{-1}(E)$). Though $C(E)$ is exponentially large near the transition
point, it depends on the energy $E$ only, so that when we keep $E$
fixed and increase the system size $N$, the number of levels in the
interval $C^{-1}(E)$ gets arbitrarily large, since the level spacing
scales as $1/N$.

Expanding both sides of Eq.(\ref{6}) over $\Psi$,
one can express $\left\langle \phi_{\sigma,b1}
\phi_{\sigma,b1}\phi_{\sigma',b1}
\phi_{\sigma',b1}\right\rangle_{g_T}$ in terms of the matrix $Q$ as
\begin{eqnarray}\nonumber
&& \left\langle \phi_{\sigma,b1}\phi_{\sigma,b1}\phi_{\sigma,b1}
\phi_{\sigma,b1}\right\rangle_{g_T}\\ \nonumber
&&= \frac{4!}{B_4}\left[\frac{1}{2}
g_{0,xx}Q_{b_1b_1}^{\sigma\sigma}Q_{b_1b_1}^{\sigma\sigma}+
 g_{0,xy}Q_{b_1b_1}^{\sigma\sigma}+g_{0,yy}\right]\ ;\\ \nonumber
&& \left\langle \phi_{R,b1}\phi_{R,b1}\phi_{A,b1}
\phi_{A,b1}\right\rangle_{g_T}\\ \nonumber
&& = -\frac{4}{B_4}\left[
g_{0,xx}\left(Q_{b_1b_1}^{RR}Q_{b_1b_1}^{AA}+
2Q_{b_1b_1}^{RA}Q_{b_1b_1}^{AR}\right)+\right. \\
&& \left.
g_{0,xy}\left(Q_{b_1b_1}^{RR}Q_{b_1b_1}^{AA}\right)+g_{0,yy}\right]\ ,
\label{7}
\end{eqnarray}
where $g_{0,xx}=\partial^2 g_0/\partial x^2|_{x,y=0}$;
$g_{0,yy}=\partial^2 g_0/\partial y^2|_{x,y=0}$;
$g_{0,xy}=\partial^2 g_0/\partial x\partial y|_{x,y=0}$,
and $B_4=\int dz h(z) z^4$.
This  allows us to represent right-hand side of Eq.(\ref{4})
in the following form:
\begin{eqnarray} \nonumber
&& 2\pi^2\left[\Delta^{-1}\pi(r,E)\delta(\omega)+\Delta^{-2}
\tilde{R}_2(\omega)\sigma(r,E,\omega)\right]\\ \nonumber
&&= -\frac{4}{B_4}g_{0xx} \mbox{Re}
\left\langle\left(Q_{b_1b_1}^{RR}Q_{b_1b_1}^{AA}+
2Q_{b_1b_1}^{RA}Q_{b_1b_1}^{AR} \right.\right. \\
&& \left.\left.-\frac{1}{2}\left[
Q_{b_1b_1}^{RR}Q_{b_1b_1}^{RR}+Q_{b_1b_1}^{AA}Q_{b_1b_1}^{AA}\right]\right)
\right\rangle_Q\ ,
\label{8}
\end{eqnarray}
where
$$
\langle ... \rangle_Q=\int dQ (...)
\exp{\left(\frac{i\pi\rho\omega N}{4}
\mbox{Str } Q \Lambda\right)}
$$
The integrals over $Q-$matrices are  the standard ones \cite{Efrev},
yielding:
\begin{eqnarray}\nonumber
&& \mbox{Re}\left\langle Q_{b_1b_1}^{RR}Q_{b_1b_1}^{AA}\right\rangle_Q=
1-2R_2^{(0)}(\omega/\Delta)\\ \nonumber
&& \left\langle Q_{b_1b_1}^{RA}Q_{b_1b_1}^{AR}\right\rangle_Q=
-\frac{2i\Delta}{\pi(\omega+i0)};\\
&& \left\langle Q_{b_1b_1}^{RR}Q_{b_1b_1}^{RR}\right\rangle_Q=
\left\langle Q_{b_1b_1}^{AA}Q_{b_1b_1}^{AA}\right\rangle_Q=1\ ,
 \label{9}
\end{eqnarray}
where $R_2^{(0)}(\omega/\Delta)$ is the level correlation function in
the Gaussian Orthogonal Ensemble.
Substituting this in Eq.(\ref{8}), we finally find
\begin{equation}\label{main}
\sigma(r,E,\omega)=
\frac{1}{3}\pi(r,E)=\frac{1}{N^2}\frac{4g_{0,xx}}
{\pi^2\rho^2 B_4}
\end{equation}

The coefficient $1/3$ in Eq.(\ref{main}) corresponds to the case of
unbroken time reversal symmetry (orthogonal ensemble). For the unitary
ensemble (broken time reversal symmetry) the same consideration yields
the coefficient $1/2$ instead, so that the general relation reads
\begin{equation}
\sigma(r,E,\omega)=\frac{\beta}{\beta+2}\pi(r,E)\ ,
\label{9a}
\end{equation}
where $\beta$ is the conventional symmetry parameter equal to $\beta=1\
(2)$ for the orthogonal (resp. unitary) ensembles.
This relation between the overlap of two different eigenfunctions
$\sigma(r,E,\omega)$
and self-overlap $\pi(r,E)$
 is valid {\it everywhere} in the phase of extended eigenstates, up to
the mobility edge $E=E_c$, provided the number of sites (the system  volume)
exceeds the correlation volume. In particular, it is valid
in the critical region $|E-E_c|\ll E_c$, where a typical eigenfunction
is very sparse and self-overlap (hence, IPR) grows like
$\exp{\left(\mbox{const} |E-E_c|^{-1/2}\right)}$\cite{FM-ema}.

Eq.(\ref{9a}) implies the following structure of eigenfunctions within
an energy interval
$\delta E=\omega<C^{-1}(E)$.
Each eigenstate can be represented as a product
$\Psi_i(r)=\psi_i(r)\Phi_E(r)$.
The function $\Phi_E(r)$ is an eigenfunction envelope of "bumps and dips"
 which is smooth
on a microscopic scale comparable with lattice constant.  It is the
same for
all eigenstates around  energy E, reflects the underlying gross
(multifractal)
spatial structure and governs the divergence of self-overlap
at critical point. In contrast, $\psi_i(r)$ is Gaussian white-noise
component fluctuating in space on the scale of lattice constant.
It fills in the "smooth" component $\Phi_E(r)$ in an individual way for
each eigenfunction, but is not critical, i.e. is not sensitive to
the vicinity
of the Anderson transition. These Gaussian fluctuations  are
responsible for the factor $\beta/(\beta+2)$ (which is the same as in
the corresponding Gaussian Ensemble) in Eq.(\ref{9a}).

As was already mentioned, this picture is valid in the energy window
$\delta E\sim C^{-1}(E)$ around the energy $E$; the number of
levels in this window being large as $\delta E/\Delta\sim
NC^{-1}(E)\gg 1$ in the thermodynamic limit $N\to\infty$. These states
form a kind of Gaussian Ensemble on a spatially non-uniform
(multifractal for $E\to E_c$) background $\Phi_E(r)$. Since the
eigenfunction correlations are described by the formula (\ref{9a}),
which has exactly the same form as in the Gaussian Ensemble, it is not
surprising that the level statistics has the WD form
everywhere in the extended phase \cite{sparse}.

Let us make a side remark here.
We believe on physical grounds that the same picture should hold for a
conventional $d$-dimensional conductor. First of all, the general
mechanism of the transition is the same in $d<\infty$ and $d=\infty$
models. Furthermore, the sparsity (multifractality) of eigenstates
near the transition point takes its extreme form for $d=\infty$ models
\cite{ldos-bethe}, so that since the strong correlations (\ref{9a}) take
place at $d=\infty$ it would be very surprising if they do not hold
at finite $d$ as well. Finally, Eq.(\ref{9a}) was proven by an
explicit calculation in the weak localization regime \cite{corr}, where
$\sigma(\bbox{r},E,\omega)={\beta\over\beta+2}\pi(\bbox{r},E)=
V^{-2}[1+\Pi(\bbox{r},\bbox{r})]$, with $V$ being the system volume and
$\Pi(\bbox{r},\bbox{r})$ the diffusion propagator.
Therefore, we believe that also for a $d$-dimensional conductor the WD
statistics applies everywhere in the delocalized phase, if the system
is large enough. Furthermore, this implies that exactly in the
critical point the level repulsion has a conventional $\omega^\beta$
form on a scale $\omega\sim\Delta$. We refer the reader to a separate
publication \cite{corr1}, where these questions are discussed in
more detail.

\section{Fluctuations of the LDOS: Critical region.}
\label{s5}

Now we are prepared to study the structure of the LDOS in the vicinity
of the localization transition. We note first that
the parameter $\alpha$ defined by eq.(\ref{e30})
is nothing else but the coupling constant of
the corresponding $\sigma$-model. In fact, to derive $\sigma$-model
rigorously one has to introduce $n\gg 1$ states per lattice site
\cite{Weg,Efe85}, 
whereas the Anderson model corresponds to $n=1$. However, the both
models have essentially equivalent critical behavior, see below. 
The
$\sigma$-model is defined by the action (we consider the unitary
symmetry case for definiteness):
\begin{equation}
S\{Q\}={\alpha\over 2}\sum_{\langle ij\rangle}\mbox{Str}Q_i Q_j +
\varepsilon\sum_i\mbox{Str} \Lambda Q_i\ ,\qquad
\label{e31}
\end{equation}
where the summation in the first term goes over the pairs of nearest
neighbor lattice sites, $Q_i=T_i^{-1}\Lambda T_i$ is a $4\times 4$
supermatrix belonging to certain coset space and satisfying the
constraint $Q^2=1$, $\varepsilon=-i\pi\nu\omega/2$,
$\nu$ is the density of states,
$\Lambda=\mbox{diag}\{1,1,-1,-1\}$, and $\mbox{Str}$ denotes the
supertrace. The reader is referred to Refs.\cite{Efrev,VWZ,my} for more
detailed information on the supersymmetry formalism.

While considering the vicinity of the Anderson
transition point we prefer to work within the $\sigma-$model formalism.
The $\sigma$-model
formulation has a considerable advantage: its
behavior is determined by a single coupling constant $\alpha$.
This should be contrasted with the original Anderson model
controlled by energy $E$ and
the whole distribution function $\gamma(\epsilon)$).
However,
the mechanism underlying
the transition and all the essential features of the
critical behavior are exactly the same for both the
$\sigma$-model and the Anderson model on the Bethe lattice, see 
 Refs.\cite{MF91,Efe-bl,Zirn-bl,gruz}.

When formulated on a
Bethe lattice the nonlinear $\sigma$-model
can be reduced to a certain non-linear
integral equation for a function $Y(Q)$ on the coset space defined as
\begin{equation}
Y(Q_1)=\int\prod_{i\ne 1} DQ_i e^{-S\{Q\}}
\label{e32}
\end{equation}
Because of the invariance properties the
 function $Y(Q)$ depends only on the two
scalar variables (eigenvalues) $1\le\lambda_1<\infty$ and $-1\le
\lambda_2\le 1$. Moreover, in the localized phase (as well as close to
the transition point in the delocalized phase) only the dependence on
the ``non-compact'' variable $\lambda_1$ is essential. In particular, in
the localized phase the integral equation
Eq.(\ref{e32}) in the limit $\eta\to 0$ takes the form
\begin{equation}
y(t)=\int_{-\infty}^\infty dt' L_\alpha(t-t') \exp\{-2e^{t'}\}
y^m(t')\ ,
\label{e33}
\end{equation}
where $t=\varepsilon\lambda_1$ and
\begin{eqnarray}
&&L_\alpha(t)=\left({\alpha\over 2\pi}\right)^{1/2}e^{t/2}e^{-\alpha\cosh
t}\left[\sinh\alpha\cosh t \right. \nonumber\\
&& \left.+\left(\cosh\alpha-{\sinh\alpha\over
2\alpha}\right)\right]
\label{e34}
\end{eqnarray}
In the localized regime ($\alpha<\alpha_c$), eq.(\ref{e33}) has a
solution in the form of a kink with the asymptotics $y(t)\simeq 1$ at
$t\to -\infty$ and $y(t)\simeq 0$ at $t\to +\infty$. In contrast, in the
delocalized phase ($\alpha<\alpha_c$), eq.(\ref{e33}) has a trivial
solution $y(t)=0$ only. The Anderson transition point $\alpha_c$ is thus
determined by a condition of the stability of the kink solution, which
has the form \cite{Efe-bl,Zirn-bl}
\begin{eqnarray}
&& m w_{\alpha_c}(1/2)=1\ ; \label{e35}\\
&& w_\alpha(\theta)=\int dt e^{-\theta t} L_\alpha(t) \label{e36}
\end{eqnarray}

For the nonlinear $\sigma$-model with orthogonal symmetry (corresponding to
the case of unbroken time
reversal invariance) the underlying structure is completely
analogous,  except for a more complicated form of
the function $L_\alpha(t)$
\cite{Efe-bl,Verb-bl}. However, all the gross properties
and the results are essentially the same as in the
unitary case.

The technically simplest (but still non-trivial) "toy" variant
of the supersymmetric nonlinear $\sigma$-model was
introduced by Zirnbauer in \cite{hyper} and called ``hyperbolic
superplane'' (HSP). When formulated on the Bethe lattice
 it has the critical behavior
completely equivalent to that of the true nonlinear
 $\sigma$-models of Anderson
localization \cite{gruz}.
In particular, the "invariant" phase (which should be
associated with the localized states regime) is described by the
same integral equation (\ref{e33}), with the kernel $L_\alpha(t)$
given by
\begin{equation}
L_\alpha(t)={1\over 2 K_{1/2}(\alpha)}\exp\left({t\over 2}-\alpha\cosh
t\right)
\label{e37}
\end{equation}

The kernel (\ref{e37}) has the same gross features as
(\ref{e34}). Moreover, we are able to show that the position
of the mobility edge in the Anderson model on the large connectivity Bethe lattice is exactly given by Eqs.(\ref{e35}), (\ref{e36}) with the same
 kernel Eq.(\ref{e37}). Derivation of this result, and more generally, 
demonstration of the equivalence of the $\sigma-$model and the Anderson model
in the regime $\alpha\ll 1$ will be published elsewhere \cite{tobe}.
Let us stress, that this is exactly the region, where the transition for 
$m\gg 1$ happens, see Eq.(\ref{e39}) below.

 Substituting eq.(\ref{e37}) into
eqs.(\ref{e35}), (\ref{e36}), one gets the following equation for the
transition point \cite{gruz}:
\begin{equation}
m{K_0(\alpha_c)\over K_{1/2}(\alpha_c)}=1\ ,
\label{e38}
\end{equation}
where $K_\nu(z)$ is the modified Bessel function. Using now
$K_{1/2}(z)=(\pi/2z)^{1/2}e^{-z}$ and the small-$z$ behavior of
$K_0(z)$, $K_0(z)\simeq -\log(z/2)$, we find
\begin{equation}
\alpha_c\simeq {\pi \over 8m^2\log^2 m}\ ,\qquad m\gg 1
\label{e39}
\end{equation}

Let us first analyze the structure of the LDOS $\rho_1(E)$ in the phase of
localized states, $\alpha<\alpha_c$. In this regime, the spectrum is
discrete, so that $\rho_1(E)$ is given by a sum of $\delta$-functions
with positions in $E=E_\alpha$ and residues
$|\langle\alpha|1\rangle|^2$. Only those eigenstates
$|\alpha\rangle$, which are localized close to the site $|1\rangle$,
give appreciable contributions to this sum. The effective number of
such states is measured by $P^{-1}(\epsilon_1)$, where $P(E)$ is the
inverse participation ratio,
\begin{equation}
P(E)=\sum_\alpha|\langle\alpha|1\rangle|^4
\simeq \nu^{-1}\left\langle
\sum_\alpha|\langle\alpha|1\rangle|^4\delta(E_\alpha-E)
\right\rangle
\label{e40}
\end{equation}
Rewriting $P(E)$ in the second form (more convenient for an analytical
study), we used the fact that the states $|\alpha\rangle$ contributing
appreciably to $\rho_1(E)$ have energies $E_\alpha$ close to
$\epsilon_1$. The latter statement follows immediately from the
results of Sec.\ref{s4}, where the average LDOS was shown to be a
narrow Lorentzian around $\epsilon_1$, see eq.(\ref{e19}).

The IPR takes values between $1$ and $1/N$ corresponding to complete
localization and delocalization of eigenfunctions respectively. In the
localized phase, $P(E)$ is finite, whereas in the delocalized one it
is proportional to $1/N$ (inverse system volume), and thus is equal to
zero in the thermodynamic limit. In a $d$-dimensional conductor, when
$E$ approaches the transition point from the phase of localized
states, $P(E)$ vanishes continuously: $P(E)\propto|E-E_c|^{\pi_2}$,
where $\pi_2$ is one of the set of the exponents determining the
multifractal structure of eigenfunctions \cite{Weg-IPR}. In contrast,
on a Bethe lattice this happens in an abrupt manner
\cite{MF91,Efe-bl,Zirn-bl,Verb-bl}: $P(E)$ has a finite limiting value $P_c$
as $E$ approaches $E_c$ from the side of localized states
($\alpha\to\alpha_c-0$ in terms of the $\sigma$-model) and then jumps
discontinuously to zero in the extended phase
($\alpha=\alpha_c+0$). Therefore, the effective number of
$\delta$-peaks in $\rho_1(E)$ in the localized phase
stays finite as we approach the Anderson transition point.

This result was obtained in Refs.\cite{MF91,Efe-bl,Zirn-bl,Verb-bl}, where
the lattice connectivity was considered as a number of order of
unity. Let us study now what happens with the limiting value $P_c$ for
large connectivity $m\gg 1$. The IPR is given by the following
expression \cite{Zirn-bl,Verb-bl}
\begin{equation}
P=2\int_{-\infty}^\infty dt e^t \exp\{-2e^t\} y^{m+1}(t)\ ,
\label{e41}
\end{equation}
where $y(t)$ is the solution of eq.(\ref{e33}). For $m\gg 1$ the
transition happens at $\alpha_c\ll 1$, see eq.(\ref{e39}). It is easy
to see that the function $L_\alpha(t)$, eq.(\ref{e37}), has in this
regime a sharp maximum around $t=\log(1/\alpha)\gg 1$ with a width of
the order of unity. Taking into account the normalization property,
$\int dt L_\alpha(t)=1$, we have thus
$L_\alpha(t)\approx\delta(t-\log(1/\alpha))$. Substituting this in
eq.(\ref{e33}), we find that the factor $\exp(-2e^{t'})$ in the
r.h.s. implies that the kink in the function $y(t)$ is located at
$t\simeq\log(1/\alpha)$. We know from the analysis of
Refs.\cite{MF91,Efe-bl,Zirn-bl,gruz,Verb-bl} that the small-$t$
behavior of $y(t)$ near
the transition point is the following:
\begin{equation}
y_c(t)\simeq 1-e^{(t-t_0)/2}\ ,\qquad t<t_0\ ,
\label{e42}
\end{equation}
where $t_0$ is a constant determining the position of the kink. We
find thus $e^{t_0}\sim 1/\alpha_c$, up to a factor of the order of
unity. Substituting this in eq.(\ref{e41}), we get
\begin{equation}
P_c\simeq 2\int_{-\infty}^\infty dt e^t
\exp\{-2e^t\}\left[1-me^{(t-t_0)/2}\right] \ ,
\label{e43}
\end{equation}
so that
\begin{eqnarray}
1-P_c&=&2me^{-t_0/2}\int dt\exp\{3t/2-2e^t\}\nonumber\\
&=&m\sqrt{{\pi\over 8}}e^{-t_0/2}\sim m \alpha_c^{1/2}\sim {1\over\log
m}
\label{e44}
\end{eqnarray}
Therefore, at $m\gg 1$ not only IPR has a non-zero limit at the
transition point but this limiting value is close to unity, i.e. to
the upper bound corresponding to
extremely localized states concentrated on a single site.
Roughly speaking, the system
undergoes a transition directly from the deeply localized phase
($P$ close to unity) to the extended phase ($P=0$).

The above results, eqs.(\ref{e42})--(\ref{e44}) can be also reproduced
by solving iteratively the self-consistency equation
(\ref{e33}). Numerical results obtained in
Refs.\cite{Zirn-bl,Verb-bl} show a clear tendency of increase of $P_c$
with $m$, in full agreement with eq.(\ref{e44}).
Therefore, when the system approaches the Anderson transition point
from the localized side, the strength function $\rho_1(E)$ has a form
of the sum of $\delta$-like peaks with almost the whole of the
spectral weight (except a small fraction of the order of $1/\log m$)
concentrated in one peak.

When the system is driven through the critical point into the
delocalized phase, this picture evolves gradually. Namely, the peaks
get broadened with a width depending on the distance to the critical
point and vanishing (in the thermodynamic limit) at the critical
point. It is clear that the width is given by the scale $C^{-1}(E)$, within
which the eigenfunctions are fully correlated (see sec.\ref{s3}). This
scale is exponentially small in the vicinity of the critical point,
see eq.(\ref{e4b}), or in terms of the $\sigma$-model coupling
constant $\alpha$ \cite{Efe-bl,Zirn-bl,gruz},
\begin{equation}
C^{-1}(\alpha)\propto\exp\left\{
-c_1\left|{\alpha-\alpha_c\over\alpha_c}\right|^{-1/2}\right\}\
;\qquad \alpha>\alpha_c
\label{e45}
\end{equation}
Such a critical behavior is a peculiar feature of the tree-like
lattices, and is replaced by a conventional power-law critical
behavior for a $d$-dimensional systems with $d<\infty$
\cite{ldos-bethe}.
Therefore, in the critical vicinity of the transition point (on the
delocalized side), the peaks in $\rho_1(E)$ have an exponentially
small width of the order of $C(\alpha)$, eq.(\ref{e45}). This is also
confirmed by the behavior (\ref{e4c}) of the IPR in this region.

Let us study now the behavior of the coefficient $c_1$ of the critical
behavior (\ref{e45}) in the limit of large connectivity $m\gg 1$.
This can be again the most easily done for the HSP model. The
condition on the transition point is given by eq.(\ref{e39}), and the
scale $C(\alpha)$ is given by \cite{gruz}
\begin{equation}
C(\alpha)\simeq e^{\pi/\eta}\ ,
\label{e47}
\end{equation}
where $\eta$ is determined from the equation
\begin{equation}
m{K_{i\eta}(\alpha)\over K_{1/2}(\alpha)}=1
\label{e48}
\end{equation}
Expanding eq.(\ref{e48}) in $\eta$ and $\alpha-\alpha_c$ and using the
asymptotic behavior of the modified Bessel function,
\begin{eqnarray}
&& K_{i\eta}(\alpha)\simeq {1\over
2}\left[\Gamma(i\eta)\left({\alpha\over 2}\right)^{-i\eta}+
\Gamma(-i\eta)\left({\alpha\over 2}\right)^{i\eta}
\right]\ ,\qquad \alpha\ll 1
\nonumber \\
&& \Gamma(i\eta)={1\over i\eta}\Gamma(1+i\eta)\simeq{1\over
i\eta}(1-i\eta \bbox{C})\ ,\qquad \eta\ll 1
\nonumber
\end{eqnarray}
($\bbox{C}$ being the Euler's constant), we reduce eq.(\ref{e48}) to
the following form:
\begin{equation}
\eta^2={6\over -\log^3(\alpha_c/2)}\left[{\sqrt{\pi/2}\over
2m\alpha_c^{3/2}}-{1\over\alpha_c}\right](\alpha-\alpha_c)
\label{e51}
\end{equation}
Using now the formula (\ref{e39}) for $\alpha_c$, we find
\begin{equation}
\eta\simeq\sqrt{6\over\pi}m(\alpha-\alpha_c)^{1/2}\ ,
\label{e52}
\end{equation}
so that according to eq.(\ref{e47}),
\begin{eqnarray}
C(\alpha)&\simeq&\exp\left\{\sqrt{{\pi^3\over 6}}{1\over
m}(\alpha-\alpha_c)^{-1/2}\right\} \nonumber\\
&=&\exp\left\{\pi\sqrt{4\over 3}\log m
\left|{\alpha-\alpha_c\over\alpha_c}\right|^{-1/2}\right\}
\label{e53}
\end{eqnarray}
Therefore, the coefficient $c_1$ in eq.(\ref{e45}) behaves as
$c_1\simeq c_2\log m$, with a factor $c_2$ of the order of unity.
Analyzing the derivation of eq.(\ref{e52}), we find that the critical
behavior in the form  (\ref{e53}), (\ref{e45}) is valid for
$\alpha-\alpha_c\ll\alpha_c$, where the fluctuations are exponentially
strong. At $\alpha-\alpha_c\sim\alpha_c$ the quantity $C(\alpha)$ as
given by eq.(\ref{e45}) is still large, $C(2\alpha_c)\simeq
e^{c_1}=m^{c_2}$, in view of $m\gg 1$. However, $C(\alpha)$ is not
exponentially large anymore and ceases to be the leading factor
determining the critical behavior of most of the relevant quantities,
such as the LDOS moments and the conductivity. For this reason, it
turns out to be difficult to extend the above consideration onto the region
$\alpha-\alpha_c\gtrsim\alpha_c$. We know, however, from Sec.\ref{s4a}
that the region of relatively strong LDOS fluctuations extends up to
$\alpha\sim 1/m^2\sim\alpha_c\log m$.

\section{Application to the problem of a quasiparticle line shape in
quantum dots}
\label{s5a}

Let us now translate the obtainter results into the context of the
problem of the one-particle excitations in a quantum dot. This can be
straightforwardly done by using the relations (\ref{e1b}), 
(\ref{e1c}),  (\ref{e1d}), and  (\ref{e30}) between the parameters of
the two problems. The localization transition then corresponds to the
energy $E_c\sim \Delta(g/\log g)^{1/2}$ \cite{agkl}. In the localized
region (excitation energy $E$ below $E_c$) the bare states forming the
basis of the Fock space mix only weakly to each otner, so that the
exact eigenstates are close to the bare ones. In particular, admixture
of many-particles states to a single particle one is weak. Therefore,
only one exact eigenstate will contribute essentially to the spectral
decomposition (\ref{e2}) of a single-particle state, see fig 1a.
In the delocalized domain ($E$ above $E_c$) exact eigenstates are
superpositions of many bare ones. This is in full analogy with a
delocalized state of the tight-binding model, which covers many
(infinitely many in the thermodynamic limit) sites of the lattice. 
As a result, there are many exact eigenstates contributing to the
strength function (\ref{e2}). The corresponding envelope is of
irregular (strongly fluctuating) shape in the intermediate (critical)
regime $E_c<E<E_c'$, with $E_c'\sim\Delta g^{1/2}$, and acquires a
Breit-Wigner form at $E\gg E_c'$. The width of this Breit-Wigner
envelope (spreading width of the one-particle state) is given by 
eq.(\ref{e20}) (Golden Rule), yielding $\Gamma\sim \Delta (E/\Delta
g)^2$. 

Finally, let us remind that what we considered throughout the paper
was a tree-like model with a {\it constant} coordination number.
In reality, however, the number of states to which a state of the
$n$-th generation is coupled, decreases with increase of $n$. Let us
briefly discuss how this is expected to modify the results. The
transition will be now smeared into a crossover, since the transition
point gets ``generation-dependent'' (see also recent preprint
\cite{silv}). Furthermore, in the constant coordination number
approximation all states on the energy shell (i.e. with energies
within the spreading width $\sim\Gamma$ around $E$) get mixed at $E\gg
E_c'$. In 
contrast, now only first few generation will get mixed under this
condition (that will be however sufficient to produce the Breit-Wigner
envelope of the spectral function (\ref{e2})). Admixture of higher
generations will require higher energies. To estimate, when the
complete mixing of the states on the energy shell happens,
we note that the density of states of the generation $n$ is equal to
$$
\nu_{2n+1}(E)={1\over n!(n+1)!(2n)!\Delta}\left({E\over\Delta}\right)^{2n},
$$
which is a direct generalization of eq.(\ref{e1}).
Maximizing this expression, 
we find that a typical many-particle state belongs to
a generation with 
$n\sim(E/2\Delta)^{1/2}$, with typical energies of quasiparticles
$\sim E/n\sim(E\Delta)^{1/2}$. The level spacing of the states to which this
one is coupled is $\sim \Delta(\Delta/E)^{3/2}$. Comparing this to the
typical value of the matrix element, $V\sim\Delta/g$, we conclude that
the full mixing of the states on the energy shell (ergodicity) will be
reached at $E>E_c''$, with $E_c''\sim\Delta g^{2/3}$. The same estimate
for the ``chaotization border''  was obtained very recently by
Jacquod and Shepelyansky \cite{jacquod}.

\section{Summary}
\label{s6}

In this paper we have studied in detail the structure of the average
and typical local density of states in a tight-binding model on a
tree-like lattice with a large branching number $m\gg 1$. In the framework of
the mapping recently suggested in Ref.\cite{agkl} this local density
of states describes the shape of the quasiparticle excitation line
in a finite Fermi system (e.g., quantum dot). We have exploited
the supersymmetry approach to the problem developed previously,
see Refs.\cite{MF91,Efe-bl,Zirn-bl,gruz,Verb-bl}, and some of the
results obtained in these papers.

The results depend on the relation between two dimensionless parameters:
the branching number $m$ and the coupling constant $\alpha\sim(V/W)^2$
($V$  being the typical magnitude of the hopping matrix element in Fock space
and $W$ the width of the distribution of random site energies
$\epsilon_i$). The relation of these parameters to those of the
original quantum dot model can be found in Sec.\ref{s2}. When $\alpha\gg 1/m$
the LDOS in a given site $i$
averaged over the distribution of all random energies of other sites
$\epsilon_j,\quad j\ne i$
has a semicircular form. In the opposite case
(relevant to the quantum dot model), $\alpha\ll 1/m$,
the {\it averaged} LDOS has a Lorentzian form with a width given
by the Golden Rule, see eq.(\ref{e20}). However, the {\it typical} LDOS is
close to its average value only for $\alpha\gtrsim 1/m^2$.
In the opposite case, $\alpha\ll 1/m^2$, the LDOS fluctuations are
strong, and the averaged LDOS is not representative, see Fig.1. 
This is related to the existence of the Anderson localization transition
 at the point $\alpha=\alpha_c\sim 1/(m^2\log^2 m)$.

On the insulating
side of the transition, $\alpha<\alpha_c$, the
LDOS is given by a discrete sum of
$\delta$-function peaks.
The effective number of such peaks is
characterized by $1/P$, $P$ being the inverse participation ratio.
It is known \cite{MF91,Efe-bl,Zirn-bl,Verb-bl} that on the Bethe
lattice the inverse participation ratio has a finite limiting value
$P_c$ when the system approaches the transition point from the localized
phase. We have shown that in the limit $m\gg 1$ this limiting value
is close to unity, $1-P_c\sim1/\log m\ll 1$, so that 
almost all the
spectral weight (\ref{e2})
(except a small part of the order of $1/\log m$) is
concentrated in a single $\delta$-peak.
Roughly speaking, the system
undergoes a transition 
directly from the deeply localized phase to the extended one.

When the system is driven through the critical point into the
 phase of extended states  the peaks
get broadened, with their width depending on the distance to the critical
point and vanishing (in the thermodynamic limit) at the critical
point. The width is determined by the scale $C^{-1}(E)$ such that for
energy separations smaller than $C^{-1}(E)$ different
eigenfunctions are fully correlated (see sec.\ref{s3}). As a result,
the width
 is exponentially small in the vicinity of the critical point,
see eqs.(\ref{e4b}), (\ref{e45}).  This is also
confirmed by the behavior (\ref{e4c}) of the inverse participation
ratio  in this region.

Our results by and large confirm the picture presented by AGKL \cite{agkl}.
We have, however, quantified many features of the problem by using
the supersymmetry approach and some of the results obtained earlier
in the framework of this method.

We disagree with AGKL concerning their statement that
in the delocalized phase the level
correlation function of the tree-like model with fixed coordination
number is not of the Wigner-Dyson
form close to the transition point because exact eigenstates
are very sparse and ``do not talk to
each other''\cite{agkl}. We have shown in Sec.\ref{s3} that
the eigenfunctions of the sparse random matrix model
which are close in energy are  strongly correlated in this regime. 
This explains the
Wigner-Dyson form of the level correlation function proven in one of our
earlier publications \cite{sparse}.

For the quantum dot problem the obtained results imply the Fock-space
delocalization of a single-particle excitation at excitation energies
$E>E_c\sim\Delta(g/\log g)^{1/2}$ and formation of regular-shaped
Breit-Wigner envelopes at $E>E_c'\sim\Delta g^{1/2}$. Taking into
account decrease of the coordination number in higher generations
allows to estimate an energy, at which the complete ergodicity on the
energy shell is restored as $E_c''\sim\Delta g^{2/3}$.  

Finally, we would like to mention that an extensive study of
statistical properties of eigenstates 
of a finite system of interacting fermions was undertaken recently
by Flambaum, Izrailev, Casati, and Gribakin \cite{FI1,FI2}. In particular,
these authors discuss crossover   
from the regime of strongly fluctuating local spectral density to that  
characterized by small fluctuations, which is similar to the questions
addressed in the present paper \cite{FI-note}.  

\section{Acknowledgments.}

 This research
was supported by  the Deutsche
Forschungsgemeinschaft within SFB 195 (A.D.M.) and SFB 237 (Y.V.F.)
The authors are especially grateful to 
P.~W\"{o}lfle for critical reading of the manuscript 
and valuable comments and 
acknowledge useful discussions with V.~Akulin, V.~Flambaum,
Y.~Gefen, F.M.~Izrailev, V.E.~Kravtsov,
and A.~M\"uller-Groeling.

\begin{figure}
\caption{Schematic view of the local density of states in various
regimes depending on the relation between the coupling constant
$\alpha$ and the connectivity $m$. The dotted line represents the
average local spectral density, which has a Lorentzian form, see
eq.(\ref{e19}). a) $\alpha$ approaches the transition point $\alpha_c$
from the the phase of localized states $\alpha<\alpha_c$. The
spectral density is given by a sum of $\delta$-functions, with almost
all the spectral weight (except a small fraction of order of $1/\log
m$)  concentrated in one $\delta$-peak; b) delocalized phase
($\alpha>\alpha_c$), critical region. The $\delta$-peaks get
broadened, with a width being exponentially small, see eq.(\ref{e45}),
for $\alpha-\alpha_c\ll\alpha_c$; c) the region of weak fluctuations
of local spectral density, $\alpha\gg 1/m^2$. }
\end{figure}


\begin{references}
\bibitem{note1} In this point we disagree with the authors of
Ref.\cite{agkl} who estimated the branching number as $m=g^3/6$.
\bibitem{note2} In fact, the numerical coefficients change rapidly
with the hierarchy level, so that one needs really large values
of $E/\Delta$, and thus large values of $g$, to justify the Bethe
lattice construction.
\bibitem{note4} Note that in the quantum dot model the situation is,
in fact, reversed: the coupling constant is fixed, $\alpha\sim 1/g^2$,
whereas the connectivity changes with energy, $m\sim E^2/\Delta^2$,
see eqs.(\ref{e1b})--(\ref{e1d}).
\end{references}
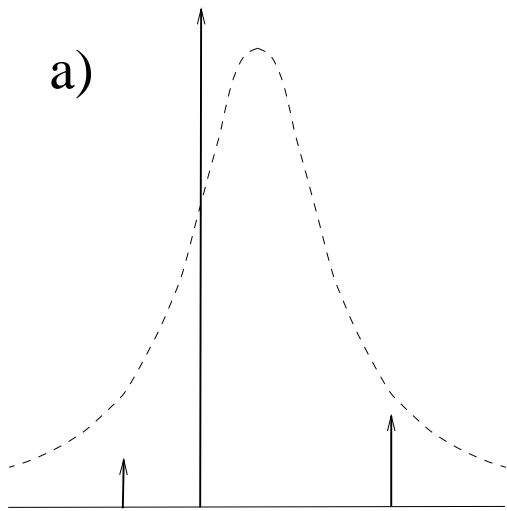
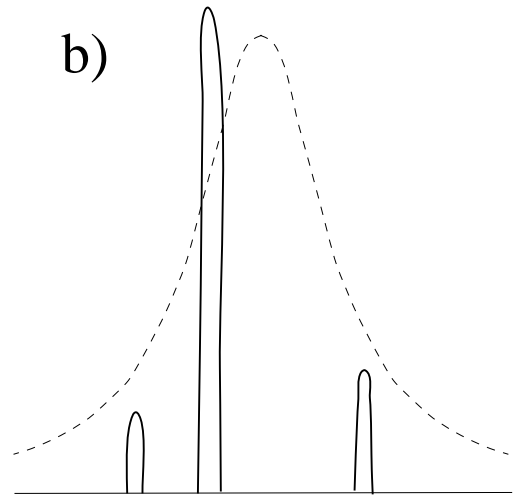
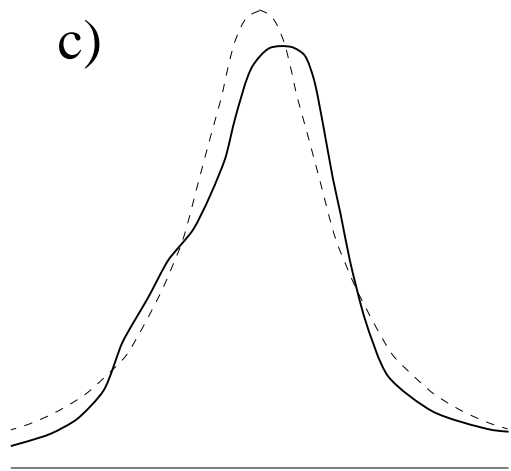
\end{document}